\documentclass[runningheads]{llncs}
\usepackage{graphicx}
\usepackage{mathtools}
\usepackage{amssymb} 

\usepackage{array}

\newcommand{\Rplus}[0]{\ensuremath{\mathbb{R}_{\geq 0}}}
\newcommand{\Real}[0]{\ensuremath{\mathbb{R}}}
\newcommand{\Nats}[0]{\ensuremath{\mathbb{N}}}

\usepackage[ruled]{algorithm2e}
\usepackage{dcolumn} 

 \SetAlFnt{\small}
\SetAlCapFnt{\small} \SetAlCapNameFnt{\small} \SetAlCapHSkip{0pt}
\IncMargin{-\parindent} \usepackage[T1]{fontenc} 
\usepackage{subfigure} \usepackage{url} \usepackage{rotating} \usepackage{color}
\def\CPP{\texttt{C++}} 

\def\argmax{\mathop{\rm argmax}} 

\def\lik{\mathcal{L}} 

\def\ef{f}

\newcommand{\comment}[1]{}

\def\rnode{\rho} 
\newcommand{\node}[1]{{\rnode \mathsf{#1}}} 
\def\inode{{\node{v}}} 
\def\dep{{\mathsf{d}_{\inode}}} 

\def\rp{s} 
\def\RP{\Sz} 

\def\rpl{m} 

\def\srp{s} 
\def\SRP{\Sz} 

\def\maxpts{\overline{\#}} 
\def\maxpsi{\overline\psi} 
\def\maxlvs{\overline{m}} 

\def\srplvplus{\Lz^{+}} 
\def\rplv{\Lz^{\bigtriangledown}} 

 \def\PQMC{\mathop{\tt PQMC}}

\makeatletter \providecommand{\leftsquigarrow}{%
  \mathrel{\mathpalette\reflect@squig\relax}%
} \newcommand{\reflect@squig}[2]{%
  \reflectbox{$\m@th#1\rightsquigarrow$}%
} \makeatother

\input{intmacros.sty}

\begin{document}
\title{Scalable Multivariate
  Histograms}
\titlerunning{Scalable Multivariate Histograms}
%
\author{
Raazesh Sainudiin\inst{1,2}
\and Warwick Tucker\inst{1,3}
\and Tilo Wiklund\inst{1,2}
}
\authorrunning{R.~Sainudiin \and W.~Tucker \and T.~Wiklund}
%
\institute{Department of Mathematics, Uppsala University, Uppsala, Sweden \and
  Combient Competence Centre for Data Engineering Sciences, Uppsala University, Uppsala, Sweden \and
  School of Mathematics, Monash University, Melbourne, Australia\\
  \email{raazesh.sainudiin@math.uu.se} \\
  \email{warwick.tucker@monash.edu}\\
  \email{tilo.wiklund@math.uu.se}}
\maketitle 
\begin{abstract}



  We give a distributed variant of an adaptive histogram estimation procedure
  previously developed by the first author. The procedure is based on regular
  pavings and is known to have numerous appealing statistical and arithmetical
  properties. The distributed version makes it possible to process data sets
  significantly bigger than previously. We provide prototype implementation
  under a permissive license.

\end{abstract}

\section{Introduction} \label{Sect:Introduction}

Suppose our random variable $X$ has an unknown density $f$ on $\Rz^d$, then for
all Borel sets $A \subseteq \Rz^d$,
$$\mu(A) := \Pr\{X \in A\}=\int_A f(x) dx \enspace . $$  
Any density estimate $f_n(x) := f_n(x;X_1,X_2,\ldots,X_n) : \Rz^d \times \left(
  \Rz^d \right)^n \to \Rz$ is a map from $\left(\Rz^d\right)^{n+1}$ to
$\Rz$. 
The objective in density estimation is to estimate the unknown $f$ from an
independent and identically distributed (IID) sample $X_1,X_2,\ldots,X_n$ drawn
from $f$. Density estimation is often the first step in many learning tasks,
including, anomaly detection, classification, regression and clustering. Current
density estimators do not computationally cope with large datasets. We use a
MapReduce implementation of an optimally smoothed multivariate density estimator
over a dense class of tree-based data-adaptive partitions. \comment{ The quality of
  $f_n$ is naturally measured by how well it performs the assigned task of
  computing the probabilities of events. Histograms and kernel density
  estimators can approximate $f$ in an asymptotic setting, i.e., as the number
  of data points $n$ approaches infinity (the so-called {\em asymptotic
    consistency} of the estimator $f_n$). But for a fixed $n$, however large but
  finite, the density estimate $f_n$ is a {\em smoothed} representation of the
  observed data \cite{Whittle1958}. Thus, a density estimator must not only be
  asymptotically $L_1$ consistent, i.e., $\int \abs(f_n(x)-f(x))dx \to 0$ as $n
  \to \infty$, but must also be optimally smoothed for any fixed sample of size
  $n$ and generalizable to unobserved data. In other words, for a given $n$ and
  a smoothing parameter $s$, our estimator should optimize a sensible expected
  objective function $g$ using a hold-out method such as cross-validation, i.e.,
  $f_{n,s^*}=\argopt_{f_{n,s}}E(g(f_{n,s},f))$.
  This asymptotically $L_1$ consistent and optimally smoothed density estimate
  $f_{n,s^*}$ of the unknown $f$ gives us a means of computing the probabilities
  of any Borel set $A \in \mathcal{B}^d$ or of computing the density at any
  point $x \in \Rz^d$. }

\subsection{Plan of the paper}
The rest of the paper is laid out as follows. Section~\ref{Sect:ArithAlgebra}
introduces the algebra for regular pavings (RPs), the arithmetic and algebra of real-mapped regular pavings ($\Rz$-MRPs) 
and statistical regular pavings (SRPs), and
explains how various multivariate data-adaptive histogram estimates can be
built using these structures. Section~\ref{Sec:RPQs:DataAdaptive} illustrates the use of a
novel partitioning strategy using a complementary pair of Priority-queued Markov chains (PQMCs) on the state
space of SRPs with a reference to a proof of its asymptotic $L_1$-consistency and a strategy for smoothing using penalized likelihoods. 
In Section~\ref{S:Distributed}, the main focus of this paper, we show how the algorithm 
can be distributed to scale to arbitrarily large datasets. In Section~\ref{S:ImpAndRes} 
we share references to the implementation in \CPP{} and Apache Spark and evaluate the performance on simulated data. 

\section{Statistical regular pavings and histograms}\label{Sect:ArithAlgebra}
This section introduces the notions of regular pavings (RPs), statistical
regular pavings (SRPs) and real mapped regular pavings ($\Rz$-MRPs), and
explains how a histogram density estimate can be built using these tree based
and arithmetically amenable data structures. 
\subsection{Regular pavings (RPs)}\label{Sect:RP}
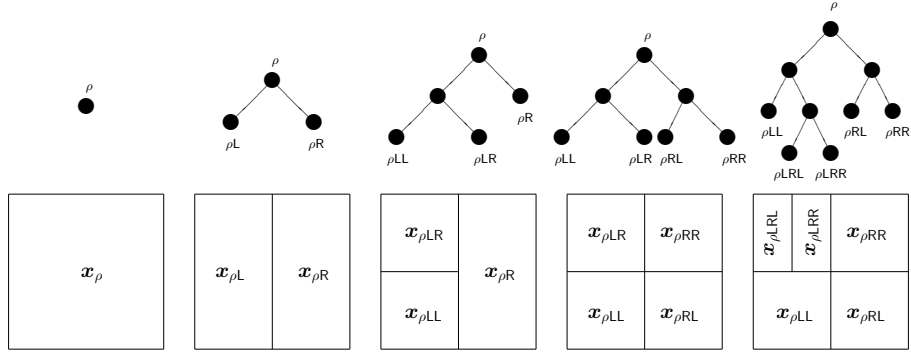
\begin{figure}[tp]
  \begin{center}
    \scalebox{0.55}{ \setlength{\unitlength}{5cm}
      \begin{picture}(6, 1.7)(0,0)
        \put(0.375, 1.175){\circle*{0.08}}
        \put(0.373,1.255){$\rho$}
        \put(0,0){\line(0,1){0.75}}  
        \put(0,0){\line(1,0){0.75}}  
        \put(0.75,0){\line(0,1){0.75}} 
        \put(0,0.75){\line(1,0){0.75}} 
        \put(0.35, 0.35){\Large{$\x_\rho$}}
        \put(1.275, 1.3){\circle*{0.08}}
        \put(1.273,1.38){$\rho$}
        \put(1.075, 1.1){\circle*{0.08}}
        \put(1.05,0.975){$\rho \sf{L}$}
        \put(1.475, 1.1){\circle*{0.08}}
        \put(1.45,0.975){$\rho \sf{R}$}
        \put(1.275, 1.3){\line(-1,-1){0.2}}
        \put(1.275, 1.3){\line(1,-1){0.2}}
        \put(0.9,0){\line(0,1){0.75}}  
        \put(0.9,0){\line(1,0){0.75}}  
        \put(1.65,0){\line(0,1){0.75}} 
        \put(0.9,0.75){\line(1,0){0.75}} 
        \put(1.275,0){\line(0,1){0.75}} 
        \put(1.0, 0.35){\Large{$\x_{\rho \mathsf{L}}$}}
        \put(1.4, 0.35){\Large{$\x_{\rho \mathsf{R}}$}}
        \put(2.275, 1.425){\circle*{0.08}} 
        \put(2.273,1.5){$\rho$}
        \put(2.275, 1.425){\line(-1,-1){0.2}}
        \put(2.075, 1.225){\circle*{0.08}} 
        \put(2.075, 1.225){\line(-1,-1){0.2}}
        \put(1.875, 1.025){\circle*{0.08}} 
        \put(1.83,0.9){$\rho \sf{LL}$}
        \put(2.075, 1.225){\line(1,-1){0.2}}
        \put(2.275, 1.025){\circle*{0.08}} 
        \put(2.25,0.9){$\rho \sf{LR}$}
        \put(2.275, 1.425){\line(1,-1){0.2}}
        \put(2.475, 1.225){\circle*{0.08}} 
        \put(2.46,1.1){$\rho \sf{R}$}
        \put(1.8,0){\line(0,1){0.75}}  
        \put(1.8,0){\line(1,0){0.75}}  
        \put(2.55,0){\line(0,1){0.75}} 
        \put(1.8,0.75){\line(1,0){0.75}} 
        \put(2.175,0){\line(0,1){0.75}} 
        \put(1.8,0.375){\line(1,0){0.375}} 
        \put(1.915, 0.545){\Large{$\x_{\rho \mathsf{LR}}$}}
        \put(1.915, 0.15){\Large{$\x_{\rho \mathsf{LL}}$}}
        \put(2.3, 0.35){\Large{$\x_{\rho \mathsf{R}}$}}
        \put(3.075, 1.425){\circle*{0.08}} 
        \put(3.073,1.5){$\rho$}
        \put(3.075, 1.425){\line(-1,-1){0.2}}
        \put(2.875, 1.225){\circle*{0.08}} 
        \put(2.875, 1.225){\line(-1,-1){0.2}}
        \put(2.675, 1.025){\circle*{0.08}} 
        \put(2.64,0.9){$\rho \mathsf{LL}$}
        \put(2.875, 1.225){\line(1,-1){0.2}}
        \put(3.075, 1.025){\circle*{0.08}} 
        \put(3.0,0.9){$\rho \sf{LR}$}
        \put(3.075, 1.425){\line(1,-1){0.2}}
        \put(3.275, 1.225){\circle*{0.08}} 
        \put(3.175, 1.025){\circle*{0.08}} 
        \put(3.15,0.9){$\rho \mathsf{RL}$}
        \put(3.275, 1.225){\line(1,-1){0.2}}
        \put(3.475, 1.025){\circle*{0.08}} 
        \put(3.45,0.9){$\rho \sf{RR}$}
        \put(3.275, 1.225){\line(-1,-2){0.1}}
        \put(2.7,0){\line(0,1){0.75}}  
        \put(2.7,0){\line(1,0){0.75}}  
        \put(3.45,0){\line(0,1){0.75}} 
        \put(2.7,0.75){\line(1,0){0.75}} 
        \put(3.075,0){\line(0,1){0.75}} 
        \put(2.7,0.375){\line(1,0){0.75}} 
        \put(2.8, 0.545){\Large{$\x_{\rho \mathsf{LR}}$}}
        \put(2.8, 0.15){\Large{$\x_{\rho \mathsf{LL}}$}}
        \put(3.15, 0.15){\Large{$\x_{\rho \mathsf{RL}}$}}
        \put(3.15, 0.545){\Large{$\x_{\rho \mathsf{RR}}$}}
        \put(3.975, 1.55){\circle*{0.08}} 
        \put(3.973,1.65){$\rho$}
        \put(3.975, 1.55){\line(-1,-1){0.2}}
        \put(3.975, 1.55){\line(1,-1){0.2}}
        \put(3.775, 1.35){\circle*{0.08}} 
        \put(3.775, 1.35){\line(-1,-2){0.1}}
        \put(3.775, 1.35){\line(1,-2){0.1}}
        \put(4.175, 1.35){\circle*{0.08}} 
        \put(4.175, 1.35){\line(-1,-2){0.1}}
        \put(4.175, 1.35){\line(1,-2){0.1}}
        \put(3.675, 1.15){\circle*{0.08}} 
        \put(3.64,1.03){$\rho \mathsf{LL}$}
        \put(3.875, 1.15){\circle*{0.08}} 
        \put(4.075, 1.15){\circle*{0.08}} 
        \put(4.04,1.03){$\rho \mathsf{RL}$}
        \put(4.275, 1.15){\circle*{0.08}} 
        \put(4.24,1.03){$\rho \mathsf{RR}$}
        \put(3.875, 1.15){\line(-1,-2){0.1}}
        \put(3.875, 1.15){\line(1,-2){0.1}}
        \put(3.775, 0.95){\circle*{0.08}} 
        \put(3.7,0.83){$\rho \mathsf{LRL}$}
        \put(3.975, 0.95){\circle*{0.08}} 
        \put(3.9,0.83){$\rho \mathsf{LRR}$}

        \put(3.6,0){\line(0,1){0.75}}  
        \put(3.6,0){\line(1,0){0.75}}  
        \put(4.35,0){\line(0,1){0.75}} 
        \put(3.6 ,0.75){\line(1,0){0.75}} 
        \put(3.975,0){\line(0,1){0.75}} 
        \put(3.6,0.375){\line(1,0){0.75}} 
        \put(3.7875,0.375){\line(0,1){0.375}} 
        \put(3.7, 0.45){\begin{rotate}{90}{\Large{$\x_{\rho \mathsf{LRL}}$}}\end{rotate}}
        \put(3.9, 0.45){\begin{rotate}{90}{\Large{$\x_{\rho \mathsf{LRR}}$}}\end{rotate}}
        \put(3.71, 0.15){\Large{$\x_{\rho \mathsf{LL}}$}}
        \put(4.05, 0.15){\Large{$\x_{\rho \mathsf{RL}}$}}
        \put(4.05, 0.545){\Large{$\x_{\rho \mathsf{RR}}$}}
      \end{picture}
    }
  \end{center}
  \caption{A sequence of selective bisections of boxes (nodes) along the first
    widest coordinate, starting from the root box (root node) in two dimensions,
    produces an RP.}
  \label{Fig:TreeSeq2}
\end{figure}

Let $\x :=[\ul{x},\ol{x}]$ be a compact real interval with lower bound $\ul{x}$
and upper bound $\ol{x}$, where $\ul{x} \leq \ol{x}$. Let the space of such
intervals be $\Iz\Rz$. The width of an interval $\x$ is $\wid(\x) :=
\ol{x}-\ul{x}$. The midpoint is $\Imid(\x) :=
\left(\ul{x}+\ol{x}\right)/2$. 
A box of dimension $d$ with coordinates in $\Delta := \{1,2,\ldots,d\}$ is an
interval vector with $\iota$ as the first coordinate of maximum width:
\begin{displaymath}
  \x := [\ul{x}_{1}, \ol{x}_{1}] \times \ldots \times [\ul{x}_{d}, \ol{x}_{d}] =: \underset{j \in \Delta}{\otimes} [\ul{x}_{j}, \ol{x}_{j}], \quad \iota := \min\left( \operatorname*{argmax}_{i} (\wid(\x_i))\right) \enspace.
\end{displaymath}
The set of all such boxes is $\Iz\Rz^d$, i.e., the set of all interval real
vectors in dimension $d$. A \emph{bisection} or \emph{split} of $\x$
perpendicularly at the mid-point along this first widest coordinate $\iota$
gives the left and right child boxes of $\x$
\begin{equation*}\label{eqn:leftchild}
  \x_{\sf L} :=  [\underline{x}_{1}, \overline{x}_{1}] \times \ldots
  \times [\underline{x}_{\iota},\Imid(\x_{\iota}))
  \times [\underline{x}_{\iota+1}, \overline{x}_{\iota+1}]
  \times \ldots \times [\underline{x}_d, \overline{x}_d] \enspace ,
\end{equation*}
\begin{equation*}\label{eqn:rightchild}
  \x_{\sf R} :=  [\underline{x}_{1}, \overline{x}_{1}] \times \ldots
  \times [\Imid(\x_{\iota}), \overline{x}_{\iota}]
  \times [\underline{x}_{\iota+1}, \overline{x}_{\iota+1}]
  \times \ldots \times [\underline{x}_d, \overline{x}_d] \enspace .
\end{equation*}
Such a bisection is said to be {\em regular}. Note that this bisection gives the
left child box a half-open interval $[\underline{x}_{\iota},\Imid(\x_{\iota}))$
on coordinate $\iota$ so that the intersection of the left and right child boxes
is empty. A recursive sequence of selective regular bisections of boxes, with
possibly open boundaries, along the first widest coordinate, starting from the
root box $\x_{\rho}$ in $\Iz\Rz^d$ is known as a {\em regular paving}
\cite{KiefferScan00} or $n$-tree \cite{Samet1990} of $\x_{\rho}$. A regular
paving of $\x_{\rho}$ can also be seen as a binary tree formed by recursively
bisecting the box $\x_{\rho}$ at the root node. Each node in the binary tree has
either no children or two children. These trees are known as plane binary trees
in enumerative combinatorics \cite[Ex.~6.19(d), p.~220]{Stanley1999} and as
finite, rooted binary trees (frb-trees) in geometric group theory
\cite[Chap.~10]{Meier2008}. The relationship of trees, labels and partitions is
illustrated
in 
Figure~\ref{Fig:TreeSeq2} via a sequence of bisections of a square
(2-dimensional) root box 
by always bisecting on the {\em first} widest coordinate.

%
Let $\mathbb{N}:=\{1,2,\ldots\}$ be the set of natural numbers. Let the $j$-th
interval of a box $\x_{\inode}$ be $[\ul{x}_{\inode,j}, \ol{x}_{\inode,j}]$, the
volume 
of a $d$-dimensional box $\x_{\inode}$
be 
$\vol(\x_{\inode}) = \prod_{j=1}^d (\ol{x}_{\inode,j} - \ul{x}_{\inode,j})$, the
set of all nodes of an RP be $\Vz := \rnode \cup \{ \rnode\{{\mathsf L},
{\mathsf R} \}^j : j \in \Nz\}$, the set of all leaf nodes be $\Lz$ and the set
of internal nodes or splits be $\breve{\Vz}(\rp) :=\Vz(\rp)\setminus\Lz(\rp)$.
The set of leaf boxes of a regular paving $\rp$ with root box $\x_{\rnode}$ is
denoted by $\x_{\Lz(\rp)}$ and it specifies a partition of the root box
$\x_{\rnode}$. Let $\RP_k$ be the set of all regular pavings with root box
$\x_{\rnode}$ made of $k$ splits. Note that the number of leaf nodes $\rpl =
\lvert \Lz(\rp) \rvert =k+1$ if $\rp \in \RP_k$.
The number of distinct binary trees with $k$ splits is equal to the Catalan
number $C_k$. 
\comment{
  \begin{equation}\label{EqCatalanNumber}
    C_k = \frac{1}{k+1}\binom{2k}{k} = \frac{(2k)!}{(k+1)!(k!)} \enspace .
  \end{equation}
} For $i,j\in \Zz_+$, where $\Zz_+ := \{0,1,2,\ldots\}$ and $i \leq j$, let
$\RP_{i:j}:=\cup_{k=i}^j \RP_{k}$ be the set of regular pavings with $k$ splits
where $k \in \{i,i+1,\ldots,j\}$. Let the set of all regular pavings be
$\RP_{0:\infty} := \lim_{j \to \infty} \RP_{0:j}$.
\subsubsection{Statistical regular pavings (SRPs)}\label{Sect:SRP}

A \textit{statistical regular paving} (SRP) denoted by $\srp$ is an extension of
the RP structure that is able to act as a partitioned `container' and responsive
summarizer for multivariate data. An SRP can be used to create a histogram of a
data set.
A recursively computable statistic \cite{Fisher1925,GrayMoore2003} that an SRP
node $\inode$ caches is $\#\x_{\inode}$, the count of the number of data points
that fell into $\x_{\inode}$.
A leaf node $\inode$ with $\#\x_{\inode} > 0$ is a non-empty leaf node. The set
of non-empty leaves of an SRP $\srp$ is $\srplvplus(\srp) := \{\inode \in
\Lz(\srp) : \#\x_{\inode} > 0\} \subseteq
\Lz(\srp)$. 

\begin{figure}[htbp]
  \centering \subfigure[{An SRP tree and its constituents.}]{ \scalebox{0.65}{
      \setlength{\unitlength}{5.0cm}
      \begin{picture}(1.55, 2.0)(1.5, -2.3) \put(2.275,
        -0.575){\circle*{0.08}} 
        \put(2.273,-0.5){$\rho$} \put(1.8,-2){\line(0,1){0.75}} 
        \put(1.8,-2){\line(1,0){0.75}} 
        \put(2.55,-2){\line(0,1){0.75}} 
        \put(1.8,-1.25){\line(1,0){0.75}} 
        \put(2.275, -0.575){\line(-1,-1){0.2}} \put(2.075,
        -0.775){\circle*{0.08}} 
        \put(2.04,-0.88){${\rho\mathsf{L}}$} \put(2.275,
        -0.575){\line(1,-1){0.2}} \put(2.475, -0.775){\circle*{0.08}} 
        \put(2.35,-0.8){${\rho}\mathsf{R}$}
        \put(2.175,-2){\line(0,1){0.75}} 
        \put(2.075, -0.775){\line(-1,-1){0.2}} \put(1.875,
        -0.975){\circle*{0.08}} 
        \put(1.83,-1.1){${\rho}\mathsf{LL}$} \put(2.075,
        -0.775){\line(1,-1){0.2}} \put(2.275, -0.975){\circle*{0.08}} 

        \thicklines
        {\color[rgb]{0.5, 0.5, 0.5}{\put(2.275,-0.975){\vector(-3,-4){0.38}}}}
        \put(1.88, -1.5){\circle*{0.02}}

        \thicklines
        {\color[rgb]{0.5, 0.5, 0.5}{\put(2.275,-0.975){\vector(-4,-3){0.45}}}}
        \put(1.82, -1.33){\circle*{0.02}} 

        \thicklines
        {\color[rgb]{0.5, 0.5, 0.5}{\put(2.275,-0.975){\vector(-1,-3){0.18}}}}
        \put(2.07, -1.55){\circle*{0.02}} 

        \thicklines
        {\color[rgb]{0.5, 0.5, 0.5}{\put(2.475, -0.775){\vector(-1,-4){0.23}}}}
        \put(2.23, -1.72){\circle*{0.02}} 

        \thicklines
        {\color[rgb]{0.5, 0.5, 0.5}{\put(2.475, -0.775){\vector(-1,-3){0.23}}}}
        \put(2.22, -1.5){\circle*{0.02}} 

        \thinlines
        {\color[rgb]{0.5, 0.5, 0.5}{\put(2.475, -0.775){\vector(0,-5){0.55}}}}
        \put(2.475, -1.35){\circle*{0.02}} 

        {\color[rgb]{0.5,0.5,0.5}
          \thinlines
          \put(2.475, -1.225){\oval(0.25,0.9)[tr]}
          \put(2.475, -1.225){\oval(0.25,0.9)[br]}
          \put(2.65, -1.1){\line(0,1){0.2}}
          \put(2.475, -1.675){\vector(-1, 0){0}}
        }
        \put(2.45, -1.675){\circle*{0.02}}

        {\color[rgb]{0.5,0.5,0.5}
          \thinlines
          \put(2.475, -1.325){\oval(0.35,1.1)[tr]}
          \put(2.35, -1.325){\oval(0.6,1.1)[br]}
          \put(2.6, -0.875){\line(0,-1){0.7}}
          \put(2.35, -1.875){\vector(-1, 0){0}}
        }
        \put(2.33, -1.875){\circle*{0.02}} 

        {\color[rgb]{0.5,0.5,0.5}
          \thinlines
          \put(1.875, -1.375){\oval(0.35,0.8)[tl]}
          \put(2.0, -1.375){\oval(0.6,0.8)[bl]}
          \put(1.7, -1.1){\line(0,-1){0.05}}
          \put(2.0, -1.775){\vector(1, 0){0}}
        }
        \put(2.02, -1.77){\circle*{0.02}} 

        {\color[rgb]{0.5,0.5,0.5}
          \thinlines
          \put(1.875, -1.425){\oval(0.25,0.9)[tl]}
          \put(1.875, -1.425){\oval(0.25,0.9)[bl]}
          \put(1.875, -1.875){\vector(1, 0){0}}
        }
        \put(1.9, -1.875){\circle*{0.02}} 

        \put(2.23,-1.1){${\rho}\mathsf{LR}$}
        \put(2.33,-0.59){{$\#\x_{\rho} = 10$}}
        \put(1.7,-0.79){{$\#\x_{\rho \sf{L}} = 5$}}
        \put(2.53,-0.79){{$\#\x_{\rho \sf{R}} = 5$}}
        \put(1.48,-0.99){{$\#\x_{\rho \sf{LL}} = 2$}}
        \put(1.9,-0.99){{$\#\x_{\rho \sf{LR}} = 3$}}
        \put(1.8,-1.625){\line(1,0){0.375}} 
        \put(1.92, -1.45){{\Large{$\x_{\rho\mathsf{LR}}$}}}
        \put(1.92, -1.85){{\Large{$\x_{\rho\mathsf{LL}}$}}}
        \put(2.28, -1.65){{\Large{$\x_{\rho\mathsf{R}}$}}}

\end{picture}
\label{Fig:CacheSRP}} }
\centering \subfigure[{An SRP histogram $f_{n,\rp}$.}]
{ \scalebox{0.65}{
    \setlength{\unitlength}{5.0cm}
    \begin{picture}(0.25,2.0)(1.1, -2.3) \put(2.275,
      -0.575){\circle*{0.08}} 
      \put(2.273,-0.5){$\rho$} \put(2.275, -0.575){\line(-1,-1){0.2}}
      \put(2.075, -0.775){\circle*{0.08}} 
      \put(2.04,-0.88){${\rho\mathsf{L}}$} \put(2.275, -0.575){\line(1,-1){0.2}}
      \put(2.475, -0.775){\circle*{0.08}} 
      \put(2.45,-0.88){${\rho}\mathsf{R}$} \put(2.075,
      -0.775){\line(-1,-1){0.2}} \put(1.875, -0.975){\circle*{0.08}} 
      \put(1.83,-1.1){${\rho}\mathsf{LL}$} \put(2.075, -0.775){\line(1,-1){0.2}}
      \put(2.275, -0.975){\circle*{0.08}} 
      \put(2.23,-1.1){${\rho}\mathsf{LR}$}
      \put(2.53,-0.79){{$\frac{5}{10} \times \frac{1}{1/2}=1$}}
      \put(1.350,-0.99){{$\frac{2}{10}\times \frac{1}{1/4}=0.8$}}
      \put(2.33,-0.99){{$\frac{3}{10}\times \frac{1}{1/4}=1.2$}}
    \end{picture}
  } \includegraphics[width=4.75cm]{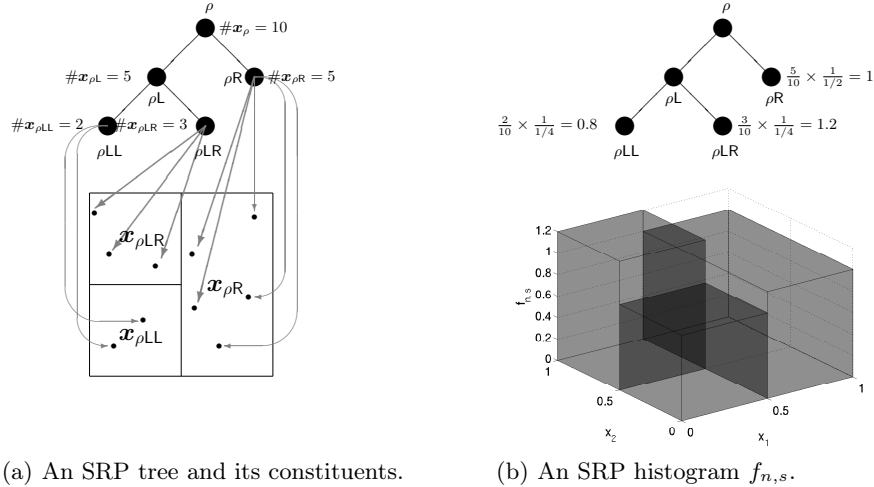}
  \label{Fig:SimpleHist}
}
\caption{{An SRP and its corresponding histogram.}\label{F:SRPTreeAndHist}}
\end{figure}

Figure~\ref{F:SRPTreeAndHist} depicts a small SRP $\srp$ with root box
$\x_{\rnode} \in \Iz\Rz^2$. The number of sample data points in the root box
$\x_{\rnode}$ is 10. Figure~\ref{Fig:CacheSRP} shows the tree, including the
count associated with each node in the tree and the partition of the root box
represented by the leaf boxes of this tree, with the sample data points
superimposed on the boxes. Figure~\ref{Fig:SimpleHist} shows how the density
estimate is computed from the count and the volume
of 
leaf boxes to obtain the density estimate $f_{n,\rp}$ as an SRP histogram.

An SRP histogram is obtained from $n$ data points that fell into $\x_{\rho}$ of
SRP $s$ as follows:
\begin{equation}\label{EqSRPHistogram}
  \ef_{n,s}(x) =
  \ef_n(x) =
  \sum_{\inode \in \Lz(\srp)} \frac{\BBs{1}_{\x_{\inode}}(x)}{n}\left(\frac{\#\x_{\inode}}{\vol(\x_{\inode})}\right)\enspace.
\end{equation}
It is the maximum likelihood estimator over the class of simple
(piecewise-constant) functions given the partition $\x_{\Lz(\srp)}$ of the root
box of $\srp$. We suppress subscripting the histogram by the SRP $s$ for
notational convenience. SRP histograms have some similarities to dyadic
histograms (for eg.~\cite[chap.~18]{KlemelaBook2009}, \cite{BSP2013}). Both are
binary tree-based and partition so that a box may only be bisected at the
mid-point of one of its coordinates, but the RP structure restricts partitioning
further by only bisecting a box on its first widest coordinate in order to make
$\RP_{0:\infty}$ closed under addition and scalar multiplication and thereby
allowing for computationally efficient computer arithmetic over a dense set of
simple functions. See \cite{MRP2012} for statistical applications of this
computer tree arithmetic, including conditional density regression, $1-\alpha$
confidence sets and associated tail probabilities, and model averaging across
multivariate histograms with different partitions.

The set of regularly paved real-valued simple functions on $\x_{\rnode} \in
\Iz\Rz^d$ satisfies the conditions of a Stone-Weierstrass theorem and is
therefore dense in $\mathcal{C}(\x_{\rnode},\Rz)$, the algebra of real-valued
continuous functions over $\x_{\rnode}$ \cite[Theorem~4.1]{MRP2012}. This
ensures that histograms obtained from statistical regular pavings subject to an
asymptotically $L_1$-consistent partitioning strategy can uniformly approximate
any continuous density $f: \x_{\rnode} \to \Rz$. 
Note that $\RP_{0:\infty}$, the set of RP tree partitions, contains the set of
dyadic partitions of $\x_{\rnode}$ represented by complete dyadic binary trees:
$\{\RP_0,\RP_2,\RP_4, \RP_8, \ldots \} \subset \RP_{0:\infty}$. Thus
$\RP_{0:\infty}$ can be used in principle to obtain any $\sigma$-additive
measure using standard measure-theoretic constructions (but such constructions
without efficiency considerations may be limited in practice due to finite
machine memory and computing time).

\comment{
  The volume is associated with the depth of a node. The depth of a node
  $\inode$ in an RP is denoted by $\dep$. A node has depth $\dep = k$ in the
  tree if it can be reached by $k$ splits from the root node. If an RP has root
  box $\x_{\rnode}$ and a node $\inode$ in the regular paving has depth $k$,
  then the volume of the box $\x_{\inode}$ associated with that node is
  $\vol(\x_{\inode}) =
  2^{-k}\vol(\x_{\rnode})$. 
  Any tree can be uniquely identified by the {\em sequence of its leaf node
    depths} if a consistent ordering of leaf nodes is used, say, lexicographic
  left-right or $\textsf{L}$-$\textsf{R}$ ordering. For example the leaf nodes
  of the regular paving in Figure~\ref{Fig:SimpleHist}, listed in left-to-right
  order, are $[\rnode \textsf{LL}, \rnode \textsf{LR}, \rnode \textsf{R}]$.
%
  Node $\rnode \textsf{R}$ has depth 1, node $\rnode \textsf{LL}$ has depth 2,
  etc. The sequence $2,2,1$ uniquely identifies the tree and the tree (as
  discussed above) uniquely identifies the partition of the root box, and so the
  sequence of leaf node depths also uniquely describes the partitioning of the
  root box $\x_{\rnode}$. In general, an RP is denoted by $\rp$. Where it is
  convenient, an RP may be labelled by its leaf node depth sequence. For
  example, the final regular paving in
  Figure~\ref{Fig:SimpleHist} can be referred to as $\rp_{2,2,1}$ or simply
  $\rp_{221}$ if the maximal depth is less than $10$.

\begin{figure}[!t]
  \centerline{
    \includegraphics[scale=0.4]{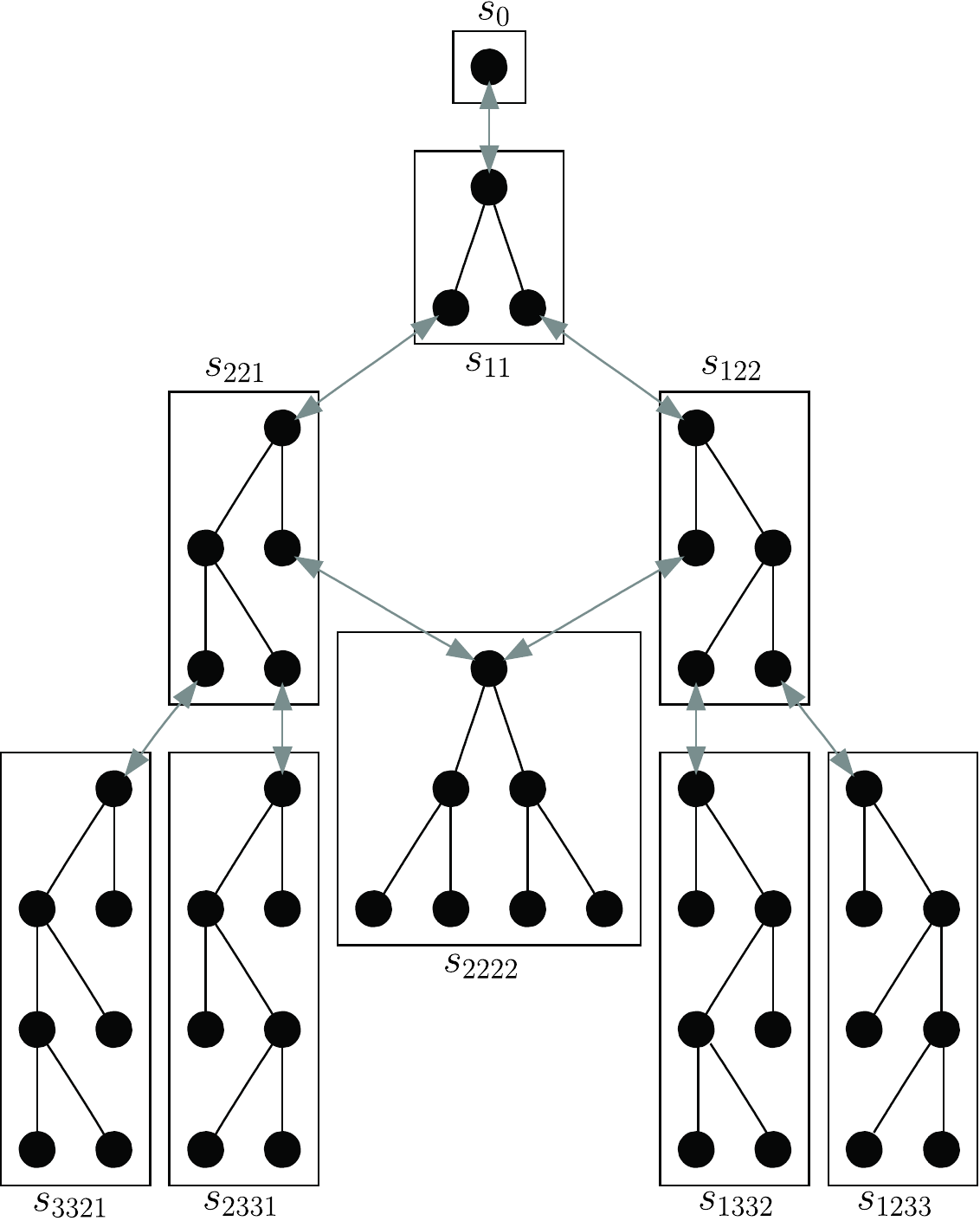}}
  \caption{Hasse (or transition) diagram over $\RP_{0:3}$ with split/reunion
    transitions from one regular paving tree to another.}
  \label{Fig:StateSpace}
\end{figure}

Figure~\ref{Fig:StateSpace} displays the Hasse or transition diagram over
$\RP_{0:3}$ where the gray arrows represent the immediate precedence relation
from one regular paving tree to another through a split or reunion. There may be
more than one path from the root node to a particular regular paving in $\RP_k$,
i.e., more than one distinct sequence of $k$ splits may result in the same
regular paving in $\RP_k$. In Figure~\ref{Fig:StateSpace}, for example, there
are two paths to $\rp_{2222}$. More generally the number of distinct paths from
the root node to the regular paving tree $s$, by splitting leaf nodes
recursively, is given by:
\begin{equation}\label{E:CatalanCoefficient}
  \mathcal{C}(\rp)  = \prod_{\inode \in \breve{\Vz}(\rp)}\binom{\wedge_{\inode}^L+\wedge_{\inode}^R}{\wedge_{\inode}^L} = \left(\lvert \Lz(\rp) \rvert -1\right)!\prod_{\inode \in \breve{\Vz}(\rp)}\frac{1}{(\wedge_{\inode}^L+\wedge_{\inode}^R+1)} \enspace,
\end{equation}
where $\wedge_{\inode}^L$ and $\wedge_{\inode}^R$ are the number of split nodes
in the left and right subtrees below the internal node $\inode \in
\breve{\Vz}(\rp)$ in the regular paving tree $\rp$ with $k=\lvert \Lz(\rp)
\rvert -1$ many splits. This product of binomial coefficients of splits,
$\mathcal{C}(s)$, is called the {\em shape functional} by
\cite[Cor.~4.1]{Dobrow1995} or the {\em Catalan coefficient}
\cite{CatalanCoeff2012,SainudiinVeber2016} and is the solution to an enumerative
combinatorial exercise \cite[Ch.~3, Ex.~1.b., p.~312]{Stanley1997}. The binomial
terms in the product can be directly understood as the number of distinct ways
in which the splits to the left and right of each internal node can be
interleaved when counting the number of distinct paths through $\RP_{0:k}$ from
the root to a given $\rp \in \RP_k$.
Randomized algorithms for data adaptive partitioning schemes here are Markov
chains on $\mathbb{S}_{0:\infty}$.



Without loss of generality, let $f$ be a probability density supported on
$\x_{\rho}=[0,1]^d$. Consider the Markov chain $\{S(t)\}_{t\in\Zz_+}$ on
$\RP_{0:\infty}$ initialized at the root regular paving $\rp(0)$ with root box
$\x_{\rho}$. 
For the current RP state $\rp$ of the chain, associate the {\em split transition
  probability} $p(\inode)$ for each leaf node $\inode \in \Lz(s)$, such that
$\sum_{\inode \in \Lz(\rp)} p(\inode)=1$.
%
If $p(\inode)$ only depends on invariant features of $\inode$, such as
$\x_{\inode}$, $\vol(\x_{\inode})$ and $\int_{\x_{\inode}}f(x)dx$, then all the
path probabilities of reaching an RP $\rp = \rp(k) \in \RP_k$, are identically
equal to:
\begin{multline*}
  \Pr\left(S(0)=\rp(0),S(1)=\rp(1),\ldots,S(k)=\rp(k)=\rp \right) \\
  = \prod_{t=1}^k \Pr\left( S(t)=s(t) \vert S(t-1)=s(t-1) \right) \Pr\left(
    S(0)=s(0) \right) = \prod_{\inode \in \breve{\Vz}(\rp)} p(\inode) \enspace ,
\end{multline*}
because the order in which one splits the internal nodes of $\rp$ in
$\breve{\Vz}(\rp)$, corresponding to each distinct path to reach $\rp$ from the
root, does not affect the total product of probabilities. Therefore, by
\eqref{E:CatalanCoefficient}, the probability that the chain visits a particular
$\rp \in \RP_k$ is simply:
\begin{equation}\label{E:primitiveStateProb}
  \Pr\left(S(k)=\rp \vert S(0)=s(0)\right) = \mathcal{C}(\rp) \prod_{\inode \in \breve{\Vz}(\rp)} p(\inode) = k! \prod_{\inode \in \breve{\Vz}(\rp)} \frac{p(\inode)}{(\wedge_{\inode}^L+\wedge_{\inode}^R+1)}
\end{equation}

Now, we get our {\em yang chain} that splits leaves proportional to the
probability of the corresponding leaf boxes with $p$ in
\eqref{E:primitiveStateProb} given by:
\begin{equation}\label{E:yangP}
  p(\inode)=\int_{\x_{\inode}}f(x)dx \enspace,
\end{equation} 
and our {\em yin chain} that proportionally splits the leaf nodes with the least
probability but with the most Lebesgue measure (or $\vol$) with $p$ in
\eqref{E:primitiveStateProb} given by:
\begin{equation}\label{E:yinP}
  p(\inode)=\frac{\tilde{p}(\inode)}{N_{\rp}}, \quad \tilde{p}(\inode) := \left( 1-\int_{\x_{\inode}}f(x)dx\right)\vol(\x_{\inode}), \quad N_{\rp}:=\sum_{\inode \in \Lz(\rp)} \tilde{p}(\inode) \enspace.
\end{equation} 
The exact probabilities for these ``yin and yang'' Markov chains can be obtained
from the above three equations: \eqref{E:primitiveStateProb}, \eqref{E:yangP}
and \eqref{E:yinP}. The yin and yang chains have their natural empirical
extensions over the space of statistical regular pavings (SRPs) such that
recursively computable statistics at each node $\inode$ determine its transition
probability $p(\inode)$ and thereby uphold \eqref{E:primitiveStateProb}. For
instance, $\int_{\x_{\inode}}f(x)dx$ is replaced by the empirical measure
$\#\x_{\inode}/n$ in \eqref{E:yangP} and \eqref{E:yinP} to obtain the empirical
variants over SRPs for the yang and yin chains, respectively. These chains and
their empirical variants were chosen as our primitive complementary Markov
chains due to their simplicity, computational ease and interpretability.

Finally, our primitive yin and yang pair of chains can be made to collaborate in
complementary ways in order to stochastically search for optimal data-adaptive
partitions by simply starting a set of yang chains from states along the path of
a yin chain (reminiscent of a tributary system of paths taken by independent
yang chains starting from states along a path taken by the yin chain). They are
complementary because the yin chain focuses on carving away large regions of the
root box with hardly any density while the yang chain initialized from various
support-carved states of the yin chain can focus on the splitting of regions
with the most density in order to achieve the classical statistical rule of
nearly {\em statistically equivalent blocks} \cite{Tukey1947}.
%
As proved in \cite{SainudiinTeng2019}, the priority-queued empirical variants of
the yin and yang chain yield data-adaptive RP partitions
that 
satisfy the three conditions of \cite{LugosiNobel1996} for their asymptotic
$L_1$ consistency.

}

\subsection{Priority-queued Markov chains}\label{Sec:RPQs:DataAdaptive}

The pseudo-code to generate sample paths from a generic priority-queued Markov
chain (PQMC) over the state space of statistical regular pavings is given in
Algorithm~\ref{A:PQMC}. A leaf node of a statistical regular paving (SRP) is
splittable if it contains data and the child nodes from the split can be
represented in the computer (as described in the next paragraph). A PQMC is a
Markov chain on SRPs whose transition probabilities are given by ordering the
elements of $\rplv(\srp)$, the splittable leaf nodes of the current SRP state
$\srp$, by a randomized queue that is prioritized according to a given priority
function $\psi: \rplv(\srp) \to \Rz$. This priority-queued collection of
splittable leaf nodes is used to select the next node to be split from
$\argmax_{\inode \in \rplv(\srp)} \psi(\inode)$, the set of splittable leaf
nodes of $\srp$ which are equally `large' when measured using $\psi$. If there
is more than one such `largest' node the choice is made uniformly at random from
this set; this is the `randomized' aspect of the process.
Three criteria can be specified to stop the PQMC. A straightforward stopping
condition is to stop partitioning when the number of leaves in the SRP reaches a
specified maximum $\maxlvs$. The other stopping condition relates to the
priority function so that partitioning stops when the value of the largest node
under the priority function $\psi$ is less than or equal to a specified value
$\maxpsi$. A PQMC will also stop partitioning if there are no splittable leaf
nodes in the
SRP. 

\begin{algorithm}
  \SetKwInOut{Input}{input} \SetKwInOut{Output}{output}
  \SetKwInOut{Initialize}{initialize} \SetKwComment{Comment}{// }{}
  \DontPrintSemicolon
  \caption{$\PQMC(\srp, \psi, \maxpsi, \maxlvs)$}
  \label{A:PQMC}
  \Input{
    $\srp$, initial SRP with root box $\x_{\rnode}$,\\
    $x = (x_1,x_2,\ldots,x_n)$, a data burst of size $n$,\\
    $\psi\,:\,\rplv(\srp) \rightarrow \Rz$, a priority function,\\
    $\maxpsi$, maximum value of $\psi(\inode) \in \rplv(\srp)$ for any
    splittable
    leaf node in the final SRP,\\
    $\maxlvs$, maximum number of leaves in the final SRP.\\
  } \Output{a sequence of SRP states $[\srp(0),\srp(1),\ldots,s(T)]$ such that
    $\rplv(\srp(T)) = \emptyset$ or $\psi(\inode) \leq \maxpsi$ $\forall\inode
    \in \rplv(\srp(T))$ or $\lvert\Lz(\srp(T))\rvert \leq \maxlvs$ .} \BlankLine
  \Initialize{
    $\x_{\rnode} \leftsquigarrow x$, make $\x_{\rnode}$ such that $\cup_i^n x_i \subset \x_{\rnode}$ if $\nexists$ domain knowledge or historical data,\\
    $\srp \leftsquigarrow \x_{\rnode}$, specify the root box of $\srp$,\\
    $\mathbf{s} \gets [\srp]$\\
  } \While{$\rplv(\srp) \neq \emptyset \And \lvert \Lz(\srp) \rvert < \maxlvs
    \And \psi\left(\argmax_{\inode \in \rplv(\srp)} \psi(\inode)\right) >
    \maxpsi$} { $\inode \gets \mathtt{random\_sample}\left(\underset{\inode \in
        \rplv(\srp)}{\argmax } \, \psi(\inode) \right)$ \Comment*{sample
      uniformly from nodes with largest $\psi$} $\srp \gets \srp$ with node
    $\inode$ split
    \Comment*{split the sampled node and update $\srp$}
    $\mathbf{s}.\mathtt{append}(\srp)$ \Comment*{append the new SRP state with
      an additional split}
  }
\end{algorithm}

The output of $\PQMC(\srp, \psi, \maxpsi, \maxlvs)$ algorithm is
$[\srp(0),\srp(1),\ldots,\srp(T)]$, a sequence of SRP states giving a sample
path from the PQMC $\{S(t)\}_{t \in \Zz_+}$ on $\SRP_{0:\maxlvs-1}$, such that
$\rplv(\srp(T)) = \emptyset$ or $\psi(\inode) \leq \maxpsi$ $\forall\inode \in
\rplv(\srp(T))$ or $\lvert \Lz(\srp(T)) \rvert \leq \maxlvs$ and $T$ is a
corresponding random stopping time. Care must be taken in defining splittable
nodes as well as the values of the stopping parameters $\maxpsi$ and $\maxlvs$,
to ensure that $\psi(\inode) \leq \maxpsi$ $\forall\inode \in \rplv(\srp(T))$
\emph{and} $\lvert \Lz(\srp(T)) \rvert \leq \maxlvs$.
%
If the initial state $S(t=0)$ is the root $\srp \in \SRP_0$ then PQMC
$\{S(t)\}_{t \in \Zz_+}$ on $\SRP_{0:\maxlvs-1}$ satisfies $S(t) \in \SRP_{t}$
for each $t \in \Zz_+$, i.e., the state at time $t$ has $t+1$ leaves or $t$
splits. Note that the initial state can be specified by a sample from any
distribution on $\SRP_{0:\maxlvs-1}$. In fact, we will use a distribution
defined from sample paths of one PQMC with a specific priority function to
initialize several independent PQMCs with a different and complementary priority
function to find our 
density estimate as described
next. 

\subsubsection{Statistically Equivalent Blocks PQMC or SEB-PQMC}

A statistically equivalent block (SEB) partition of a sample space is some
partitioning scheme that results in equal numbers of data points in each element
(block) of the partition \cite{Tukey1947} (except possibly in blocks on the
boundary of the partitioned space). A statistically equivalent blocks
(SEB)-based SRP partitioning scheme specified by the PQMC with (i) priority
function $\psi(\inode)=\#\x_{\inode}$, i.e., the number of sample points
associated with a node $\inode$, (ii) $\psi$-related stopping condition
$\maxpsi=\maxpts$ and (iii) $\maxlvs$, is denoted by SEB-PQMC.
%
Thus, at stopping time $T$, the SRP $\srp$ realized by the SEB-PQMC will be such
that either $\rplv(\srp) = \emptyset$ or $\lvert \Lz(\srp) \rvert \leq \maxlvs$
or $\#\x_{\inode} \leq \maxpts$ $\forall \inode \in \rplv(\srp)$. The operation
may only be considered to be successful if $\lvert \Lz(\srp) \rvert \leq
\maxlvs$ and $\#\x_{\inode} \leq \maxpts$ $\forall \inode \in \rplv(\srp)$. Care
must be taken to ensure that the operation is successful.
Therefore, an SEB-PQMC can be used to create a final SRP at stopping time $T$
such that each leaf node has at most $\maxpts$ of the sample data points
associated with it and the total number of leaves is at most $\maxlvs$.
%
Intuitively, SEB-PQMC prioritizes the splitting of leaf nodes with the largest
numbers of data points associated with them.

\begin{figure}[htpb]
  \centering \subfigure[20 leaves.]{
    \includegraphics[scale=0.18]{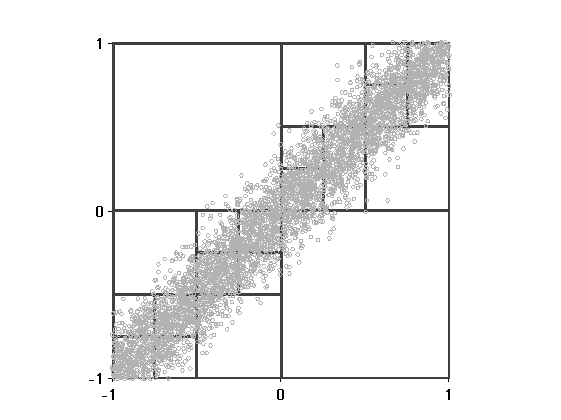}
    \label{Subfig:CarverSEBEffect_SEB1}
  }%
  \subfigure[40 leaves.]{
    \includegraphics[scale=0.18]{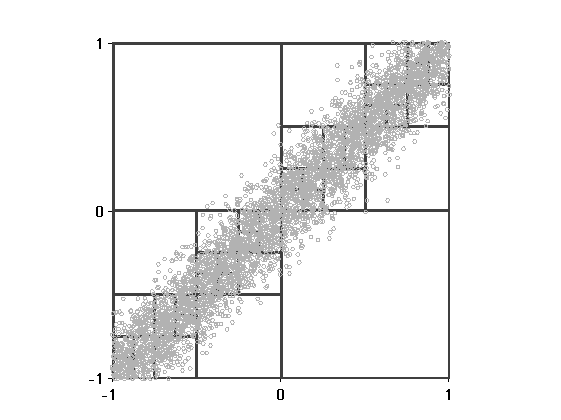}
    \label{Subfig:CarverSEBEffect_SEB2}
  } \subfigure[20 leaves.]{ \includegraphics[scale=0.18]
    {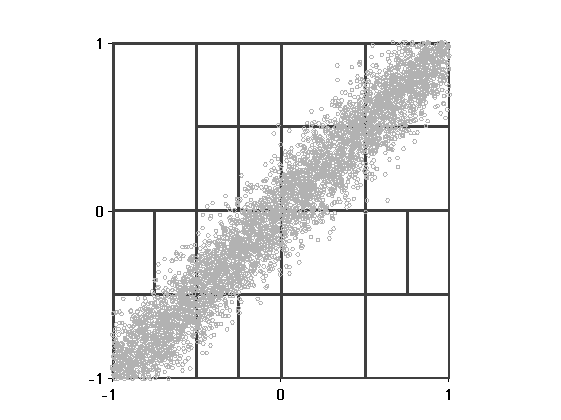}
    \label{Subfig:CarverSEBEffect_Carving1}
  }%
  \subfigure[40 leaves.]{ \includegraphics[scale=0.18]
    {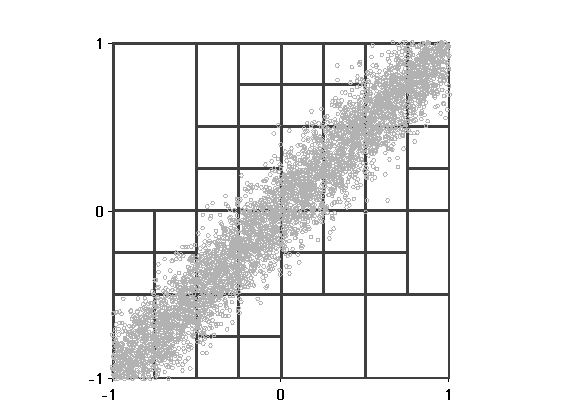}
    \label{Subfig:CarverSEBEffect_Carving2}
  }
  \caption{Partition using an SEB-PQMC and
    SPC-PQMC.}\label{Fig:CarverSEBEffect_SEB}
\end{figure}

Unfortunately, under an SEB-PQMC partitioning strategy over SRPs, the nodes with
least data associated with them will remain unsplit for longer (and will
possibly never be split). This tends to result in relatively large regions of
very low density in the tails of the density estimate $\ef_n$ formed from the
SRP. Figure~\ref{Fig:CarverSEBEffect_SEB} shows two partitions of an SRP
associated with a highly correlated dataset super-imposed on it. As the number
of leaves in the partition increases from 20
(Figure~\ref{Subfig:CarverSEBEffect_SEB1}) to 40
(Figure~\ref{Subfig:CarverSEBEffect_SEB2}), large sub-boxes containing very
little sample data remain unsplit. The effect can be especially distorting to
the resulting histogram when the axis-aligned root box required by the SRP is a
poor fit to the data (when large areas of the root box contain no data points).
Thus, we would like to carve out empty space within the root box while remaining
in $\SRP_{0:\infty}$. The next PQMC ameliorates this undesirable feature of
SEB-PQMC and can be used collaboratively with SEB-PQMC to still ensure
asymptotic $L_1$ consistency of the joint partitioning strategy.


\subsubsection{Support Carving PQMC or SPC-PQMC}\label{Sect:Carver}
A support-carving (SPC) SRP partitioning scheme specified by the PQMC with (i)
$\psi(\inode)=\Xi(\inode)=(1-\#\x_{\inode}/n)\vol(\inode)$, the SPC priority
function which prioritizes node $\inode$ according to the relative lack of its
empirical measure when further scaled by its volume, (ii) $\psi$-related
stopping condition $\ol{\Xi}$ and (iii) maximum number of leaves
$\maxlvs^{\Xi}$, is denoted by SPC-PQMC.
Using SPC-PQMC to carve out ``empty space'' in the complement of the empirical
support can be thought of as an `inversion' of the SEB-PQMC:
instead of prioritizing splitting of nodes with the largest number of data
points associated with them as done by SEB-PQMC, the SPC-PQMC prioritizes
splitting of non-empty leaf nodes with large boxes but few data points and thus
is likely to result in one of the child nodes being devoid of any sample data.
Hence, the two PQMCs are thought to be complementary.
The carving effect is illustrated in
Figure~\ref{Fig:CarverSEBEffect_SEB}. 
The partitions are created for the same set of correlated data shown in
Figure~\ref{Fig:CarverSEBEffect_SEB} using SPC-PQMC.
Figures~\ref{Subfig:CarverSEBEffect_Carving1} and
\ref{Subfig:CarverSEBEffect_Carving2} show the partition when the SRP has 20 and
40 leaves, respectively. Partitioning is concentrated in the regions of sparse
sample data and the effect is to reduce the size of the sub-boxes of the
partition into which these sparse data points fall, in effect more tightly
enclosing the support of the data as desired while still remaining in
$\SRP_{0:\infty}$.

\comment{
  \begin{figure}[htpb]
    \centering \subfigure[20 leaves.]{ \includegraphics[scale=0.40]
      {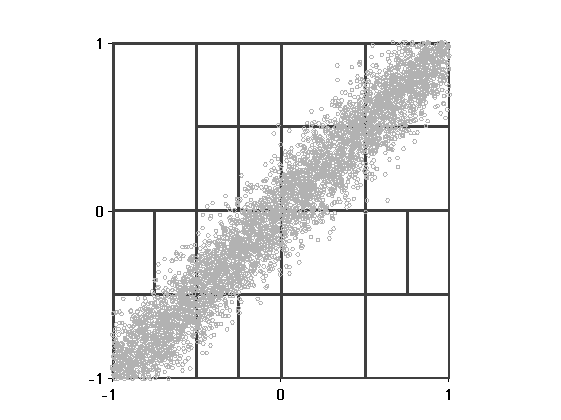}
      \label{Subfig:CarverSEBEffect_Carving1}
    }%
    \subfigure[40 leaves.]{ \includegraphics[scale=0.40]
      {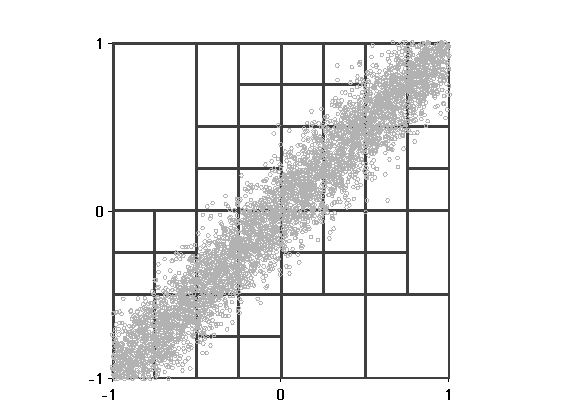}
      \label{Subfig:CarverSEBEffect_Carving2}
    }
    \caption{Partition using an SPC-PQMC.}\label{Fig:CarverSEBEffect_Carving}
  \end{figure}
}

Algorithm~\ref{A:PQMC} with the SPC priority function $\Xi$ can be used to
obtain a {\em core support-carved path} in $\SRP_{0:\maxlvs^{\Xi}}$ by
initializing from the root SRP $\srp \in \SRP_0$ through the procedure
$\PQMC(\srp, \Xi, \ol{\Xi}, \maxlvs^{\Xi})$ with $\ol{\Xi} = 0.0$. Thus, the
partitioning process of this core SPC-PQMC only stops when the SRP has
$\maxlvs^{\Xi}$ leaves or aborts if there a no splittable nodes. More general
support-carved paths can be generated by specifying $\ol{\Xi} > 0.0$ or imposing
constraints on the minimum volume of splittable nodes if deemed appropriate for
the underlying data generation process.

\subsubsection{Joint exploration}\label{S:JointExploration}
An SPC-PQMC of Section~\ref{Sect:Carver} alone will not give an effective
data-driven partitioning strategy, but used in conjunction with an SEB-PQMC it
can improve the SRP histogram. An initial SPC-PQMC can be run for a short time
(specifying $\ol{\Xi} = 0.0$ and a relatively low value of $\maxlvs^{\Xi}$, say
a small fraction of the total number of leaves $\maxlvs$), followed by an
SEB-PQMC. The empty elements of the partition `carved out' will be ignored by
the SEB-PQMC, under which partitioning will be concentrated on the areas where
most of the sample data has fallen.
The core support-carved path from SPC-PQMC over $\SRP_{0:\maxlvs^{\Xi}}$ through
$\PQMC(\srp, \Xi, \ol{\Xi}, \maxlvs^{\Xi})$ with $\ol{\Xi} = 0.0$ will be used
to determine $c^{\Xi}$ many spread-out initial conditions for launching multiple
independent SEB-PQMCs. It is along these $c^{\Xi}$ many joint SPC/SEB-PQMC
paths, which may be viewed as a system of $c^{\Xi}$ many SEB-PQMC tributaries
spread along the core support-carved path of the SPC-PQMC, that we conduct
\textsc{MAP} estimation for a given $\tau$-specific prior by simply identifying
the SRP state $\srp$ with the highest log-posterior value. The search for the
density 
estimate is conducted sequentially along each of the SPC/SEB-PQMC
paths. 
%
%
The $c^{\Xi}$ many initial conditions from the core support-carved path for
launching the SEB-PQMCs, should be well spread out in order to facilitate a
better search strategy over $\Sz_{0:\maxlvs}$, 
and in particular, contain the SEB-PQMC launched from the root node as one of
its joint SPC/SEB-PQMC paths. In essence, we can view the joint exploration as
one that will necessarily improve upon the 
estimate found by the SEB-PQMC initialized from the root node (which satisfies
the three conditions of \cite{LugosiNobel1996} and thus asymptotically
$L_1$-consistent as shown in \cite{SainudiinTeng2019}). 
\subsubsection{Smoothing}
\begin{figure}[!t]
  \centering \includegraphics[scale=0.3]{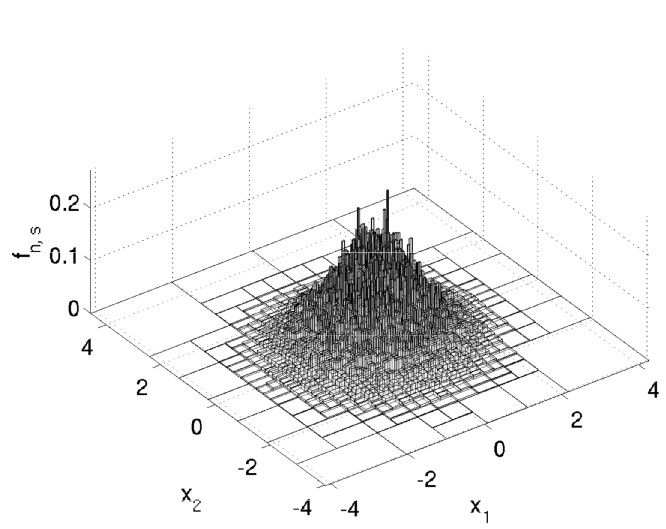}
  \includegraphics[scale=0.3]{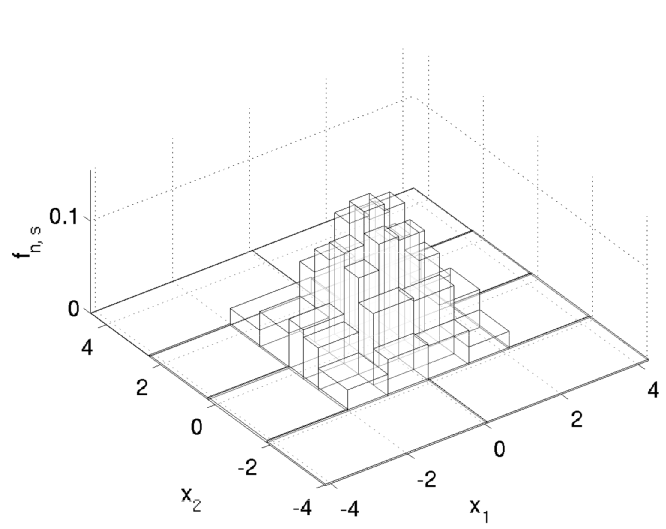}
  \caption{Two histogram density estimates for the standard bivariate Gaussian
    density. The left figure shows a histogram with 1485 leaf nodes where
    $\maxpts=50$ and the histogram on the right has $\maxpts = 1500$ resulting
    in 104 leaf nodes.}
  \label{Fig:DifferentK}
\end{figure}

Figure~\ref{Fig:DifferentK} shows two different SRP histograms constructed using
two different values of $\maxpts$ for the same dataset of $n=10^5$ points
simulated under the standard bivariate Gaussian density. A small $\maxpts$
produces a histogram that is under-smoothed with unnecessary spikes
(Fig.~\ref{Fig:DifferentK} left) while the other histogram with a larger
$\maxpts$ used as the SEB stopping criterion is over-smoothed
(Fig.~\ref{Fig:DifferentK} right). We will use the leave-one-out
cross-validation to choose an optimally smoothed
density. 

Let $x_1, \ldots, x_n$ be the sample data,
$\lik\left(\ef_n\right)$ be the likelihood of the data given SRP $\srp$,
$m=m(\srp)=\lvert \Lz(\srp) \rvert$ be the number of cells or leaf boxes in the
partition given by the SRP $\srp$, i.e., $m=k+1$ if $\srp \in \SRP_k$.
%
For a given $\tau$, finding
the 
regularized (penalized) maximum likelihood estimate, amounts to solving the
following maximization problem over $\SRP_{0:\infty}$:
\[
  \ef_{n,\tau} := \argmax_{s \in \SRP_{0:\infty}} \log (\pi(\srp; \tau)) =
  \argmax_{s \in \SRP_{0:\infty}} \ \log \lik\left(\ef_n\right) -
  {\left(\frac{m(s)}{\tau}\right)} \enspace .
\]

%
We will solve this smoothing 
problem by minimizing Stone's leave-one-out cross-validation score
$\widehat{J}(\tau)$, a nearly unbiased estimator of the expected $L_2$ loss
$\int(\ef_{n,\tau}(x)-f(x))^2dx$ (eg.~\cite{Wasserman2003}), given by:
\[
  \widehat{J}(\tau) = \int \left(\ef_{n,\tau}(x)\right)^2 dx - \frac{2}{n}
  \sum_{i=1}^n \ef_{n,\tau}^{(-i)}(x_i) \enspace,
\]
where $\ef_{n,\tau}^{(-i)}$ is the 
estimate obtained from the data set by leaving out $x_i$.
\comment{ The first term for $\widehat{J}(\tau)$ is straightforward to computer
  and the second term simplifies to the following sum
  involving 
  $\ef_{n,\tau}$ over the non-empty leaf nodes of $s$:
  \[
    -2/(n-1) \sum_{\inode \in \srplvplus(\srp)} (\#\x_{\inode} -1)
    {\ef_{n,\tau}}(\inode) \enspace .
  \]

  Thus the 
  estimate $\ef_{n,\check\tau}$ is optimally $L_2$-smoothed with the
  optimal 
  parameter:
  \[
    \check\tau = \argmin_{\tau \in (0,\infty)} \widehat{J}(\tau) \enspace .
  \]
  The actual minimization over the prior parameter $\tau \in (0,\infty)$ is
  performed by an adaptive grid-based heuristic search method within
  an 
  interval $[\ul{\tau},\ol{\tau}]$. }
%

\section{Going distributed}\label{S:Distributed}


Simply computing multiple histograms in parallel has a number of limitations.
First of all the number of histograms to compute will typically be relatively
small. More importantly one is still limited by the memory capacity of each
individual machine, since each histogram would be based on all data points.

For simplicity we will consider only axis parallel splits through cell centres,
with respect to a priority depending on volume and count (this covers support
carving and SEB PQMCs). The idea generalizes to priority functions that depend on any
recursively computable statistic and splitting hyper-planes that may depend on
some appropriate function of both the cell and the data points it contains.

Recall that Algorithm
~\ref{A:PQMC} proceeded roughly as
follows: given a sequence \(x = (x_{1}, \dotsc, x_{N})\) of (data) points in
\(\Real^{d}\) produce \(s(0), s(1), \dotsc\) the sequence of partitions given by
splitting according to some priority \(\psi \colon \Nats \times \Rplus \to
\Rplus\) starting at any \(s(0)\). Thus \(s(n+1)\) equals \(s(n)\) with one cell
split, according to the highest priority assigned by \(\psi\). Schematically it
thus proceeds as in Algorithm~\ref{TW:SketchForward}.

\begin{algorithm}
  \SetKwInOut{Input}{input} \SetKwInOut{Output}{output}
  \SetKwInOut{Initialize}{initialize} \DontPrintSemicolon
  \caption{High level sketch of sequential splitting procedures}
  \label{TW:SketchForward}
  \Input{
    {\color[gray]{0.40}\(\psi \colon \text{Count} \times \text{Volume} \rightarrow \Rplus\), a priority function,\\
      \(x_{1}, \dotsc, x_{N}\), (data) points in \(\Real^{d}\),} \\
    \(s(0)\), initial SRP, \\
    \(\maxpsi\), maximum priority for any cell node in final output }
  \Output{Increasing (refining) \(s(0), \dotsc, s(K)\), cells of \(s(K)\) have
    priority below \(\maxpsi\)}
  \Initialize{ \(r \gets s(0)\) } \While{\(r\) has any cell of priority above
    \(\maxpsi\)} {
    \textbf{output} \(r\) \\
    \(r \gets \text{\(r\) with the cell of \emph{highest priority} split}\) }
  \textbf{output} \(r\)
\end{algorithm}

This is inherently sequential; modulo special properties of the priority
functions one needs to split the current most high priority cell before being
able to determine which cell to split next. One can make the procedure parallel
by using two basic observations. Firstly, from \(s(n)\) one can (again
sequentially) reconstruct the path \(s(n), s(n-1), \dotsc, s(0)\) without
knowing \(x\). Following the path backward entails a sequence of mergers of
sibling leafs as opposed to a sequence of splits. To get the final path one
merges in order of \emph{least} priority of the parent.

\begin{algorithm}
  \SetKwInOut{Input}{input} \SetKwInOut{Output}{output}
  \SetKwInOut{Initialize}{initialize} \DontPrintSemicolon
  \caption{Backtracking from an intermediate SRP}
  \label{TW:SketchBacktrack}
  \Input{
    {\color[gray]{0.4}\(\psi \colon \text{Count} \times \text{Volume} \rightarrow \Rplus\), a priority function,}\\
    \(s(0)\), ``intermediate'' SRP \\
  } \Output{Decreasing (coarsening) \(s(0), s(1), \dotsc, s(K)\), with \(s(K)\)
    trivial}
  \Initialize{ \(r \gets s(0)\) } \While{\(r\) has more than one cell} {
    \textbf{output} \(r\) \\
    \(\mathbf{p} \gets \text{parents of leaf nodes in \(r\)}\) \\
    \(\mathbf{w} \gets \text{priority (\(\psi\)) of all \(\mathbf{p}\) based on volume/count of children}\) \\
    \(r \gets \text{\(r\) with children of the \emph{least priority} node
      according \(\mathbf{W}\) merged }\) } \textbf{output} \(r\)
\end{algorithm}

\begin{figure}
  \centering
  \begin{tabular}{m{0.25\textwidth}m{0.21\textwidth}m{0.20\textwidth}m{0.20\textwidth}m{0.07\textwidth}}
    \includegraphics[page=1]{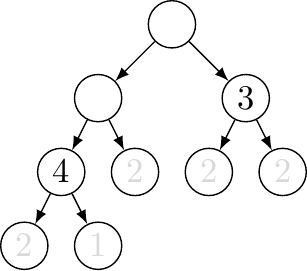} &
                                           \includegraphics[page=2]{twpic2.pdf} &
                                                                                  \includegraphics[page=3]{twpic2.pdf} &
                                                                                                                         \includegraphics[page=4]{twpic2.pdf} &
                                                                                                                                                                \includegraphics[page=5]{twpic2.pdf} \\
  \end{tabular}
  \caption[illustration of sequential and parallel procedures]{Illustration of
    Algorithm~\ref{TW:SketchBacktrack}, each tree is an SRP with nodes labelled
    with its priority and priorities of inner nodes computed recursively}
  \label{fig:twpic2}
\end{figure}

Both count and volume, and thus the priority \(\psi\), for any internal node can
be computed by adding the counts and volumes of its children. Knowing only leaf
counts and volumes one can therefore determine the last node to have been split.
This is summarized in Algorithm~\ref{TW:SketchBacktrack} and illustrated in
Figure~\ref{fig:twpic2}.

The second observation is that given a threshold \(c > 0\) the tree \(S\) given
by starting at \(s(0)\) and splitting leafs in an arbitrary order until all have
a priority less than \(c\) will satisfy \(S = s(k)\) for some \(k\). This is
most easily seen by letting \(S' = s(l)\) where \(l = \min \{ k \mid \text{all
  cells in \(s(k)\) priority below \(c\)} \}\) and noting that \(S = S'\). The
last statement holds since any cell split to get \(S\) will sooner or later have
to be split when constructing \(S'\), and vice versa. Formally this follows by
induction on the depth, within the tree, of the cell.

This gives us the freedom to pick the order in which cells are split. In
particular we can split multiple cells at once. Using the first observation the
sequence of intermediary trees can be reconstructed. While this reconstruction
is again sequential, performing mergers will generally be faster than splitting,
since the latter involves traversals of all data points in the cell.

\begin{figure}[t]
  \centering
  \begin{tabular}{m{0.06\textwidth}m{0.19\textwidth}m{0.19\textwidth}m{0.21\textwidth}m{0.22\textwidth}}
    \includegraphics[page=1]{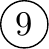} &
                                           \includegraphics[page=2]{twpic1.pdf} &
                                                                                  \includegraphics[page=3]{twpic1.pdf} &
                                                                                                                         \includegraphics[page=5]{twpic1.pdf} &
                                                                                                                                                                \includegraphics[page=6]{twpic1.pdf} \\
    \includegraphics[page=1]{twpic1.pdf} &
                                           \includegraphics[page=2]{twpic1.pdf} &
                                                                                  \multicolumn{2}{m{0.34\textwidth}}{\centering \includegraphics[page=4]{twpic1.pdf}} &
                                                                                                                                                                        \includegraphics[page=6]{twpic1.pdf} \\
  \end{tabular}
  \caption[illustration of sequential and parallel procedures]{Illustration of
    sequential (upper) and parallel (lower) splitting procedures, each tree
    represents an SRP with each node labeled by the priority of that node}
  \label{fig:twpic1}
\end{figure}

Once we can split cells simultaneously one may as well split \emph{all} nodes
with a priority over the threshold \(c\). Note that any intermediate SRP
computed this way need not lie along the path output by the original procedure;
a node may well split into two where at least one has a higher priority than
something split simultaneously (see Figure~\ref{fig:twpic1}).

The path produced by Algorithm~\ref{TW:SketchForward} can now be found using
Algorithm~\ref{TW:SketchParallel} followed by
Algorithm~\ref{TW:SketchBacktrack}. This should yield a gain in performance when
the number of data points is large, so that a merge is significantly faster than
a split, assuming one can perform all splits in each iteration in parallel.

\begin{algorithm}
  \SetKwInOut{InputOutput}{input/output}
  \SetKwInOut{Initialize}{initialize} \DontPrintSemicolon
  \caption{High level sketch of parallel splitting procedures}
  \label{TW:SketchParallel}
  \InputOutput{ As in Algorithm~\ref{TW:SketchForward}
  } \BlankLine \Initialize{ \(r \gets s(0)\) } \While{\(r\) has any cell of
    priority above \(\maxpsi\)} { \(r \gets \text{\(r\) with \emph{all} cells of
      priority above \(\maxpsi\) split}\) } \textbf{output} \(r\)
\end{algorithm}

In order for the procedure outlined above to translate into a distributed
algorithm one needs a distributed representation that allows for efficient
distributed and parallel computation of how a given set of cells split.

\begin{figure}
  \centering \includegraphics[]{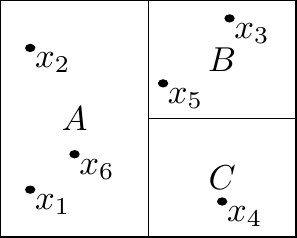}
  \caption{Partitioned points represented by \((A, x_{1}), (A, x_{2}), (B,
    x_{3}), (C, x_{4}), (B, x_{5}), (A, x_{6})\)}
  \label{fig:twpic3}
\end{figure}

The way we achieve this is by storing the current tree implicitly by tagging
each data point by the cell in which it lies. That is to say, the points
\((x_{1}, \dotsc, x_{6})\) and partition \(\{A, B, C\}\) in
Figure~\ref{fig:twpic3} would be represented by the sequence \((A, x_{1}), (A,
x_{2}), (B, x_{3}), (C, x_{4}), (B, x_{5}), (A, x_{6})\).

Counting the current number of points in each cell is now a simple reduce
operation available in, for example, Apache Spark under the name
\texttt{countByKey}.
The resulting map (counting the number of points for each cell) can either be
collected to a single machine or itself remain distributed. It involves
communication proportional to the current number of non-empty cells; in the
worst case scenario every machine has to send its counts for each non-empty
cell.

As volume is a function only of the cell this gives enough information for
determining which cells needs to be split. Simply compute the priority (a
function of count and volume) and filter to get only those cells with a priority
exceeding the predetermined threshold \(c\).

Once one has determined which cells to be split getting the new partition is a
point-wise operation. For each pair \((A, x)\) (point \(x\) in cell \(A\)) one
checks if \(A\) is to be split and if so on which side of the splitting hyper
plane \(x\) lies. If \(x\) lies below produce the pair \((A', x)\) where \(A'\)
is the lower part (relative to the hyper plane) of \(A\) and otherwise produce
\((A'', x)\) where \(A''\) is the upper part of \(A\).

Most of this computation fits directly into the MapReduce framework since it
simply applies a function point-wise (thus a map operation). How to communicate
of cells to be split depends on how the resulting map was stored; either one
broadcasts it from a single machine or performs a distributed join operation.

\begin{algorithm}
  \SetKwInOut{Input}{input} \SetKwInOut{Output}{output}
  \SetKwInOut{Initialize}{initialize} \SetKwComment{Comment}{// }{}
  \DontPrintSemicolon
  \caption{MapReduce-friendly parallel splitting procedure}
  \label{TW:SketchMapReduce}
  \Input{
    {\color[gray]{0.40}\(\psi \colon \text{Count} \times \text{Volume} \rightarrow \Rplus\), a priority function,}\\
    \(x_{1}, \dotsc, x_{N}\), (data) points in \(\Real^{d}\), \\
    \(\maxpsi\), maximum priority for any cell node in final output }
  \Output{SRP with cells of priority below \(\maxpsi\) and cell count
    corresponding to the number of points in \(x_{1}, \dotsc, x_{N}\) in the
    cell}
  \Initialize{
    \(\rho \gets \text{root cell}\) \\
    \(y \gets [(\rho, x_{1}), \dotsc, (\rho, x_{N})] \) \Comment{distributed array} \\
    \(c \gets [(\rho, N)]\) } \While{\(c\) has any cell/count pair with priority
    above \(\maxpsi\)} { \Comment{Implemented as a distributed map operation}
    \ForEach{\((a, x) \in y\)}{ \If{\(\psi(\text{count of \(a\) in \(c\)},
        \text{volume of \(a\)}) > \maxpsi\)} { \If{\(x\) is below splitting
          hyperplane of \(a\)}{ \(a' \gets \text{subcell of \(a\) below
            splitting hyperplane}\) } \Else{ \(a' \gets \text{subcell of \(a\)
            above splitting hyperplane}\) } replace \((a, x)\) by \((a', x)\) }
    } \Comment{Implemented as a distributed reduce operation}
    \(c \gets [\,]\) \\
    \ForEach{\((a, x) \in y\)}{ increment count in \(c\) at \(a\) by \(1\)
      (adding key \(a\) if needed) } } \textbf{output} SRP with leaves/counts
  corresponding to cells/counts in \(c\)
\end{algorithm}


The procedure can be improved by pruning pairs corresponding to cells with a
priority below the threshold. The relevant computations are already performed
when determining, for each tuple, whether or not the cell is one to be split
(see Algorithm~\ref{TW:SketchMapReduce}). Filtering these points avoids
repeating this check in subsequent iterations; the threshold remains fixed so
these cells will not be split in a future iteration either.

\begin{algorithm}
  \SetKwInOut{Input}{input} \SetKwInOut{Output}{output}
  \SetKwInOut{Initialize}{initialize} \SetKwComment{Comment}{// }{}
  \DontPrintSemicolon
  \caption{Optimization of Algorithm~\ref{TW:SketchParallel}}
  \label{TW:OptimFilter}
  \While{\(c\) has any cell/count pair with priority above \(\maxpsi\)} {
    \Comment{Now becomes a map and filter operation} \ForEach{\((a, x) \in y\)}{
      \If{\(\psi(\text{count of \(a\) in \(c\)}, \text{volume of \(a\)}) >
        \maxpsi\)} {
        compute new cell \(a'\) as before \\
        replace \((a, x)\) by \((a', x)\) } \Else { delete \((a, x)\) } } update
    \(c\) as before, keeping counts for cells that have passed threshold }
  \textbf{output} SRP with leaves/counts corresponding to cells/counts in \(c\)
\end{algorithm}

For good performance one needs also a good representation for the cells in the
distributed sequence of cell/point tuples. For our purposes a convenient
representation is given by identifying cells with nodes in the (infinite) plane
binary tree as described in Section~\ref{Sect:ArithAlgebra} and illustrated in
Figure~\ref{Fig:TreeSeq2}.

For practical purposes such a node can be identified with a positive natural
number as follows: the root node is assigned number \(1\), the left sub-child of
the node with number \(n\) is assigned the number \(2n\), and the right child of
the node with number \(n\) is assigned number \(2n+1\). This is simply a binary
encoding given by an initial \(1\) followed by the unique path from the root
node; left steps are encoded by \(0\) and right steps are encoded by \(1\).

Using multiple precision integer types (as provided by for example
gmp\cite{granlundgmp}) this is yields a fairly compact representation (bits
proportional to tree depth) where many operations can be implemented in terms of
bit-level operations. For example, passing to ancestors corresponds to taking
right shifts and the depth (distance to the root node) is the index of the most
significant bit.

\section{Implementation and Results}\label{S:ImpAndRes}

A complete, though non-parallel, \CPP{} implementation of regular paving trees
is available in the open source library {\tt mrs2} \cite{MRS2}. Our reference
Scala implementation of the parallel density estimation procedure on top of
Apache Spark is also publicly available \cite{SparkDensityTree}. 

Motivated by an industrial problem in fraud detection, we simulated $n=13\times
10^7$ points in $2$ and $10$ dimensions from a multivariate density. A typical
computation on an Apache Spark cluster with 5 Workers (totaling 30.0 GB Memory
and 4.4 Cores) in 2 and 10 dimensions took about 1.79 and 5.8
hours 
with 
estimated L1 errors 
of 0.03
and $1.13$, respectively. 
Density estimates for such large datasets cannot be obtained in a single
commodity machine.

\section*{Acknowledgements}
This work was initiated with support from project CORCON: Correctness by Construction, Seventh Framework Programme of the European Union, Marie Curie Actions-People, International Research Staff Exchange Scheme (IRSES) with counter-part funding from the Royal Society of New Zealand. It is completed with support from Combient Competence Centre for Data Engineering Sciences. RS thanks Nina Nowak and Guillermo Padres for inspiring discussions on the industrial needs for distributed multivariate density estimators.

\bibliographystyle{splncs04} \bibliography{references}

\begin{thebibliography}{10}
\providecommand{\url}[1]{\texttt{#1}}
\providecommand{\urlprefix}{URL }
\providecommand{\doi}[1]{https://doi.org/#1}

\bibitem{Fisher1925}
Fisher, R.A.: Theory of statistical estimation. Mathematical Proceedings of the
  Cambridge Philosophical Society  \textbf{22},  700--725 (1925)

\bibitem{granlundgmp}
Granlund, T., {the GMP development team}: {GNU MP}: {T}he {GNU} {M}ultiple
  {P}recision {A}rithmetic {L}ibrary, \url{http://gmplib.org/}

\bibitem{GrayMoore2003}
Gray, A.G., Moore, A.W.: {Nonparametric Density Estimation: Towards
  Computational Tractability}. In: SIAM International Conference on Data
  Mining. pp. 203--211. SIAM, San Francisco, California, USA (2003)

\bibitem{MRP2012}
Harlow, J., Sainudiin, R., Tucker, W.: Mapped regular pavings. Reliable
  Computing  \textbf{16},  252--282 (2012)

\bibitem{KiefferScan00}
Kieffer, M., Jaulin, L., Braems, I., Walter, E.: Guaranteed set computation
  with subpavings. In: Kraemer, W., Gudenberg, J. (eds.) Scientific Computing,
  Validated Numerics, Interval Methods, Proceedings of {SCAN} 2000, pp.
  167--178. Kluwer Academic Publishers, New York (2001)

\bibitem{KlemelaBook2009}
Klemel\"{a}, J.: Smoothing of Multivariate Data: {D}ensity Estimation and
  Visualization. Wiley, Chichester, United Kingdom (2009)

\bibitem{BSP2013}
Lu, L., Jiang, H., Wong, W.H.: Multivariate density estimation by bayesian
  sequential partitioning. Journal of the American Statistical Association
  \textbf{108}(504),  1402--1410 (2013)

\bibitem{LugosiNobel1996}
Lugosi, G., Nobel, A.: {Consistency of Data-Driven Histogram Methods for
  Density Estimation and Classification}. The Annals of Statistics
  \textbf{24}(2),  687--706 (1996)

\bibitem{Meier2008}
Meier, J.: Groups, Graphs and Trees: An Introduction to the Geometry of
  Infinite Groups. Cambridge University Press, Cambridge, United Kingdom (2008)

\bibitem{SainudiinTeng2019}
Sainudiin, R., Teng, G.: Minimum distance histograms with universal performance
  guarantees. Japanese Journal of Statistics and Data Science  \textbf{2}(2),
  507--527 (2019), \url{https://doi.org/10.1007/s42081-019-00054-y}

\bibitem{MRS2}
Sainudiin, R., York, T., Harlow, J., Teng, G., Tucker, W., George, D.: {MRS
  2.0, a C++ class library for statistical set processing and computer-aided
  proofs in statistics}. \url{https://github.com/raazesh-sainudiin/mrs2}
  (2008--2018)

\bibitem{Samet1990}
Samet, H.: The Design and Analysis of Spatial Data Structures. Addison-Wesley
  Longman, Boston (1990)

\bibitem{Stanley1999}
Stanley, R.P.: Enumerative combinatorics. {V}ol. 2, Cambridge Studies in
  Advanced Mathematics, vol.~62. Cambridge University Press, Cambridge (1999)

\bibitem{Tukey1947}
Tukey, J.W.: {Non-Parametric Estimation II. Statistically Equivalent Blocks and
  Tolerance Regions --- The Continuous Case}. The Annals of Mathematical
  Statistics  \textbf{18}(4),  529--539 (1947)

\bibitem{Wasserman2003}
Wasserman, L.: All of Statistics: A Concise Course in Statistical Inference.
  Springer, New York (2003)

\bibitem{SparkDensityTree}
Wiklund, T.: {SparkDensityTree}.
  \url{https://github.com/TiloWiklund/SparkDensityTree} (2018)

\end{thebibliography}




\end{document}